\documentclass[prd,aps,twocolumn,nofootinbib]{revtex4}

\usepackage{graphicx,amsmath}

\begin{document}

\title{MeV gamma rays from Q-ball decay}

\author{Shinta Kasuya$^a$, Masahiro Kawasaki$^{b,c}$, Naomi Tsuji$^{a,d}$}

\affiliation{
$^a$ Physics Division, Faculty of Science,
         Kanagawa University, Kanagawa 221-8686, Japan\\
$^b$ Institute for Cosmic Ray Research, the University of Tokyo, Chiba 277-8582, Japan\\
$^c$ Kavli Institute for the Physics and Mathematics of the Universe (WPI), 
        Todai Institutes for Advanced Study, the University of Tokyo, Chiba 277-8582, Japan\\
$^d$ Interdisciplinary Theoretical and Mathematical Science Program (iTHEMS), RIKEN, Saitama 351-0198, Japan}
     

\begin{abstract}
We study the supersymmetric Q balls which decay at present and find that they create a distinctive spectrum of
gamma rays at around O(10) MeV. The charge of the Q ball is lepton numbers in order for the lifetime to be 
as long as the present age of the universe, and the main decay products are light leptons. However, as the 
charge of the Q ball decreases, the decay channel into pions becomes kinematically allowed towards the end of 
the decay, and the pions are produced at rest. Immediately, $\pi^0$ decays into two photons with the energy of 
67.5~MeV, half the pion mass, which exhibits a unique emission line. In addition, $\pi^\pm$ decay into $\mu^\pm$, 
which further decay with emitting internal bremsstrahlung, whose spectrum has a sharp cutoff at $\sim$50~MeV. 
If the observations would find these peculiar features of the gamma-ray spectrum in the future, it could be 
a smoking gun of the supersymmetric Q-ball decay at present.
\end{abstract}

\maketitle

\section{Introduction}
There exist non-topological solitons, Q balls \cite{Coleman:1985ki}, which consist of scalar fields such as 
squarks and sleptons in supersymmetric theories \cite{Kusenko:1997zq}. 
Q balls with large charge can form after inflation \cite{Kusenko:1997si, Kasuya:1999wu} 
\`a la Affleck-Dine \cite{Affleck:1984fy}. Large Q balls with $Q$ being the baryon number are stable against the 
decay into nucleons, and could be dark matter of the universe \cite{Kusenko:1997si}.
Such Q ball dark matter may be detected in the large volume detectors 
\cite{Kusenko:1997vp,Arafune:2000yv,Kasuya:2001hg,Kasuya:2015uka}. 
If detected, it is not only the detection of the dark matter, but also might be the observational clue
of the supersymmetry (SUSY).

On the other hand, large Q balls with lepton charge is not stable against the decay into light leptons, 
but they could long lived. Their decay leaves some observational consequences such that they may 
result in the source of the 511~keV line gamma rays from the Galactic Center \cite{Kasuya:2005ay}, 
provide late-time entropy production \cite{Kasuya:2007cy}, or enhance the secondary gravitational waves 
\cite{Kasuya:2022cko,Kawasaki:2023rfx}.

Here we study yet another way to find the observational trace of SUSY Q balls with $Q$ being 
the lepton number. We look for clues of the Q balls whose lifetime is just as long as the present age 
of the universe. Since the charge of the Q ball (lepton number in this case) should be very large 
for such lifetime, the mass per unit charge is a little bit larger than the electron (positron) mass. Therefore,
Q balls decay into neutrinos, anti-neutrinos, electrons, and positrons in the first place, decreasing the Q-ball charge. 
As the charge gets smaller, the mass per unit charge becomes larger, and finally the decay channels 
to pions open up towards the end of the decay, and hence the pions are produced almost at rest. 
This is the unique feature of the Q-ball decay. 

Produced pions decay immediately. Neutral pions decay into two photons with the energy 
of half the pion mass, which shows a 67.5~MeV emission line, while charged pions decay into muons, 
which further decay emitting photons through internal bremsstrahlung (IB). The resulting spectrum of 
the photons has such peculiar features in the 1--100~MeV range.

The structure of the paper is as follows. In the next section, we provide basic characters of the Q ball considered in
this paper. In Sec.III, we investigate the process of the decreasing charge $Q$ for the Q-ball decay at present, and
consider the decay products and their fractions in Sec.~IV. In Sec.V, we estimate the gamma-ray spectra
produced by the Q-ball decay for both the Galactic and extragalactic emissions. Finally we conclude in Sec.~VI. 
We explain the internal bremsstrahlung in pion and muon decays in Appendix A, and the upper bound on Q-ball 
abundance in Appendix B.

\section{Q balls in the gauge mediation }
The Q ball is the energy minimum configuration of the scalar fields for the fixed charge $Q$ \cite{Coleman:1985ki}.
Those scalar fields $\Phi$, called flat directions, are all classified in terms of gauge-invariant monomials 
in the minimal supersymmetric standard model \cite{Gherghetta:1995dv,Dine:1995kz}. 
In the gauge-mediated SUSY breaking scenario, the scalar potential is written as
\begin{equation}
V(\Phi)=M_F^4 \left(\log\frac{|\Phi|^2}{M_S^2}\right)^2+m_{3/2}^2\left(1+K\log\frac{|\Phi|^2}{M_{\rm P}^2}\right)|\Phi|^2,
\label{pot}
\end{equation}
where $m_{3/2}$ is the gravitino mass and $M_{\rm P}=2.4\times 10^{18}$~GeV is the Planck mass. 
The first term comes from the gauge-mediation effects above the messenger scale $M_S$ \cite{deGouvea:1997afu}.
$M_F$ is related to the $F$ component of a gauge-singlet chiral multiplet in the messenger sector, and its range is 
given by \cite{Kasuya:2015uka}
\begin{equation}
4\times 10^4~{\rm GeV} \lesssim M_F \lesssim 0.1 \left(m_{3/2}M_{\rm P}\right)^{1/2}. 
\label{MF}
\end{equation}
The second term is due to the gravity-mediation effects. Here, $K$ is a coefficient of the one-loop corrections to the mass, 
and negative for most of the flat directions with $|K|=$0.01--0.1. Since the gravitino mass is relatively small in the 
gauge-mediation models, the second term dominates over the first one for field values larger than 
\begin{equation}
\phi_{\rm eq} \simeq \frac{\sqrt{2}M_F^2}{m_{3/2}},
\end{equation}
where we define $\Phi=\frac{1}{\sqrt{2}}\phi e^{i\theta}$.

There are two types of the Q ball depending on which term of the potential is dominant when it forms.
If the first term of Eq.(\ref{pot}) dominates the potential, the gauge-mediation type Q balls are created,
while the so-called new-type Q balls are produced when the second term overwhelms the first \cite{Kasuya:2000sc}.
The important property which concerns with following discussion is the mass per unit charge of the Q ball, $\omega_Q$, 
equivalent to the effective mass of $\Phi$ inside the Q ball:
\begin{equation}
\omega_Q \simeq \left\{
\begin{array}{ll}
\sqrt{2}\pi\zeta M_F Q^{-1/4} & ({\rm gauge\mbox{\scriptsize -}med. \ type}), \\[2mm]
m_{3/2} & ({\rm new \ type}),
\end{array}\right.
\label{omega}
\end{equation}
where $\zeta \simeq 2.5$ \cite{Hisano:2001dr}.

As shown later, we need very large charge $Q$ so that the new-type Q balls form for the field amplitude 
well above $\phi_{eq}$. The charge of the Q ball is thus estimated as \cite{Kasuya:2000wx}
\begin{equation}
Q=\beta_N\left(\frac{\phi_0}{m_{3/2}}\right)^2,
\label{form}
\end{equation}
where $\beta_N\simeq 0.01$ \cite{Hiramatsu:2010dx}, and $\phi_0$ is the amplitude of $\Phi$ at the beginning of its 
oscillations. The mass and size of this type of the Q ball are given respectively by \cite{Enqvist:1998en, Kasuya:2000sc}
\begin{equation}
M_Q \simeq m_{3/2} Q, \quad R_Q \simeq |K|^{-1/2} m_{3/2}^{-1}.
\label{masssize}
\end{equation}

\section{Q-ball decay}
Q-ball decay occurs if some decay particles carry the same kind of the charge of the Q ball and the mass of all  the 
decay particles is less than the mass of the Q ball per unit charge $\omega_Q$. Since the decay products carrying 
lepton numbers are light fermions, once the Fermi sea is filled, further decay proceeds only when produced fermions
escape from the surface of the Q ball. The upper bound of the decay rate is thus determined by the maximum 
outgoing flow of the fermions \cite{Cohen:1986ct}. This saturation occurs when the field value is large inside the
Q ball, which is the case here. Then the decay rate is estimated as \cite{Cohen:1986ct}
\begin{equation}
\Gamma_Q \simeq \frac{1}{Q}\frac{\omega_Q^3}{192\pi^2}4\pi R_Q^2 
\simeq \frac{m_{3/2}}{48\pi |K|Q},
\end{equation}
where we use Eq.(\ref{masssize}) in the last equality. Since we consider the Q balls that decay at present, 
its lifetime is set to be $\tau_Q = \Gamma_Q^{-1} \simeq t_0 \simeq 13.8$~Gyr, which leads to 
the charge of the Q ball as
\begin{equation}
Q  \simeq 2.2\times 10^{38} \left(\frac{|K|}{0.02}\right)^{-1}\left(\frac{m_{3/2}}{\rm MeV}\right).
\label{decay}
\end{equation}
From Eqs.(\ref{form}) and (\ref{decay}), we obtain the charge of the formed new-type Q ball and the gravitino mass
respectively as
\begin{equation}
Q_{\rm D} \simeq 6.3 \times 10^{38} \left(\frac{\beta_N}{0.01}\right)^{1/3}
\left(\frac{|K|}{0.02}\right)^{-2/3}\left(\frac{\phi_0}{0.3 M_{\rm P}}\right)^{2/3}, 
\end{equation}
\begin{equation}
m_{3/2} \simeq 2.9~{\rm MeV} \left(\frac{\beta_N}{0.01}\right)^{1/3}
\left(\frac{|K|}{0.02}\right)^{1/3}\left(\frac{\phi_0}{0.3 M_{\rm P}}\right)^{2/3}.
\end{equation}
Notice that the scenario works for broader parameter space, but here we take $\phi_0=0.3M_{\rm P}$ and 
$|K|=0.02$ as a typical example.\footnote{
For the gravitino mass smaller than 100~keV, the gauge-mediation type Q balls can form and decay at present to 
produce same signatures of MeV gamma rays.}

As the charge of the Q ball decreases, the new-type Q ball transforms into the gauge-mediation type 
\cite{Kasuya:2009by} when the charge becomes
\begin{equation}
Q_{\rm tr} \simeq 4\left(\frac{M_F}{m_{3/2}}\right)^4 
\simeq 2.8 \times 10^{38} \left(\frac{m_{3/2}}{2.9~{\rm MeV}}\right)^{-2},
\label{trans}
\end{equation}
where we use the upper bound of $M_F$ in Eq.(\ref{MF}) in the second equality. Thereafter, $\omega_Q$ 
increases as the charge diminishes as in the upper line in Eq.(\ref{omega}). Since pions consist of quark 
and anti-quark pairs, the decay channel to pions opens when $\omega_Q=m_\pi/2$, where the charge of 
the Q ball becomes
\begin{equation}
Q_\pi \simeq \left(\frac{2\sqrt{2}\pi\zeta}{m_\pi}\right)^4M_F^4
\simeq 3.0\times 10^{36} \left(\frac{m_{3/2}}{2.9~{\rm MeV}}\right)^2.
\label{chpi}
\end{equation}
Here the upper bound of $M_F$ in Eq.(\ref{MF}) is used in the last equality. As the decay into pions has 
just come to be kinematically allowed at that time, pions are produced at rest. This is the unique property 
of the Q-ball decay. 

\begin{figure}[ht!]
\includegraphics[width=90mm]{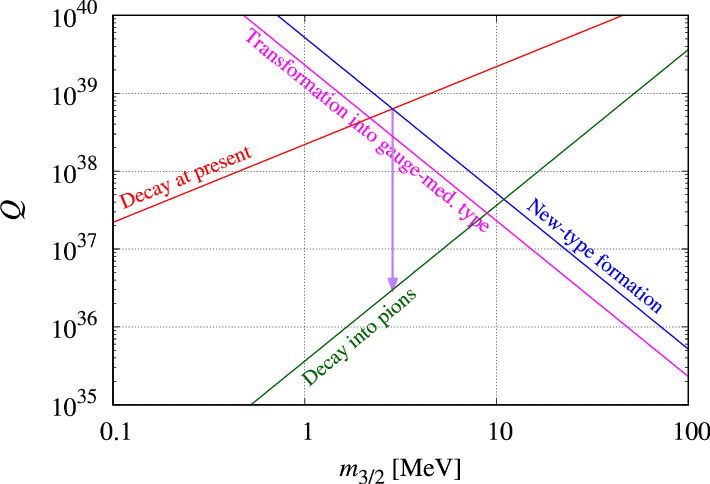} 
\caption{Evolution of the Q-ball charge is shown in the purple arrow. We show the charge at the formation (\ref{form}), 
that for the decay at present (\ref{decay}), that when the transformation from the new-type into the gauge-mediation 
type takes place (\ref{trans}), and that when the decay channel to pions opens (\ref{chpi}) as the function of $m_{3/2}$, 
for $\phi_0=0.3M_{\rm P}$ and $|K|=0.02$, in blue, red, magenta, and green lines, respectively.
\label{fig1}}
\end{figure}

In Fig.~\ref{fig1}, we show the evolution of the charge of the Q ball by the purple arrow. Here we plot the 
charge at the formation (\ref{form}), that for the decay at present (\ref{decay}), that when the transformation 
from the new-type into the gauge-mediation type takes place (\ref{trans}), and that when the decay channel to 
pions opens (\ref{chpi}) as the function of $m_{3/2}$, for $\phi_0=0.3M_{\rm P}$ and $|K|=0.02$.

\section{Decay products}
We must identify the particles emitted by the Q-ball decay. To this end, we choose the $\Phi^5={\cal Q}u{\cal Q}ue$ direction
as a concrete example below.\footnote{
Similar argument holds in other leptonic directions.}
 ${\cal Q}$ represents SU(2)$_L$-doublet squarks, $u$ and $e$ are 
SU(2)$_L$-singlet squarks and sleptons, respectively. We further assume that they consist only of the fields in 
the first generation, so that it can be written as $\tilde{u}_L\tilde{u}_R^*\tilde{d}_L\tilde{u}_R^*\tilde{e}_R^*$.
Since this direction carries lepton number $L=-1$, the field $\Phi$ possesses $L=-1/5$.

In the process of the Q-ball decay, all the decay particles must have masses smaller than $\omega_Q$, and 
some should carry anti-lepton numbers. In addition, the electric charge neutrality holds through the decay
process, since the Q ball consists of the flat direction which is gauge invariant. Thus the composition changes 
before and after the decay channel into pions opens. Before that time, it is kinematically allowed to decay 
only into electrons, positrons, neutrinos, and anti-neutrinos.\footnote{
The branching ratio to photons can be extremely suppressed due to the very small left-right mixings for
squarks and sleptons for the first and second generations \cite{Martin:1997ns}.}
Selectrons $\tilde{e}_R^*$ carrying anti-lepton number decay into anti-electron neutrinos $\bar{\nu}_e$.
Since the Q-ball decay is saturated, there is no phase space left for $\bar{\nu}_e$ to be further produced.
Therefore, the other decay particles are the following six patterns of combinations: 
$\nu_\mu\bar{\nu}_\mu\nu_\mu\bar{\nu}_\mu$, $\nu_\mu\bar{\nu}_\mu\nu_\tau\bar{\nu}_\tau$, 
$\nu_\tau\bar{\nu}_\tau\nu_\tau\bar{\nu}_\tau$, $\nu_\mu\bar{\nu}_\mu e^+ e^-$, 
$\nu_\tau\bar{\nu}_\tau e^+ e^-$, and $e^+ e^- e^+ e^-$.

On the other hand, after the decay into pions are kinematically allowed, anti-lepton numbers could be conveyed by
$e^+$, in addition to $\bar{\nu}_e$, since $\pi^-$ exists in the decay particles to compensate the electric charge of positrons.
The phase spaces of produced $\bar{\nu}_e$ and $e^+$ are filled by the saturated decay so that they cannot be
included in the remaining decay particles. In the case of $\bar{\nu}_e$, the combinations of other decay particles are
$\nu_\mu\bar{\nu}_\mu\nu_\mu\bar{\nu}_\mu$, $\nu_\mu\bar{\nu}_\mu\nu_\tau\bar{\nu}_\tau$, 
$\nu_\tau\bar{\nu}_\tau\nu_\tau\bar{\nu}_\tau$, $\pi^+\nu_e e^-$, $\pi^+\nu_\mu \mu^-$, $\pi^0 \nu_\mu\bar{\nu}_\mu$, 
$\pi^0 \nu_\tau\bar{\nu}_\tau$, $\pi^+\pi^-$, or $\pi^0\pi^0$, while they are
$\pi^-\pi^0$, $\pi^-\nu_\mu\bar{\nu}_\mu$, or $\pi^-\nu_\tau\bar{\nu}_\tau$ for $e^+$.

Now we are ready to estimate the fraction of the produced particle per unit anti-lepton number or per unit 
charge of the Q ball. After pions are kinematically allowed to produced from the Q-ball decay, the
numbers of particles per unit Q-ball charge are obtained as $f_{\pi^0}=1/12$ for $\pi^0$, $f_{\pi^+}=1/20$ 
for $\pi^+$, $f_{\pi^-}=1/15$ for $\pi^-$, $f_{\mu^-}=1/60$ for $\mu^-$, while $f_{e^+}=2/15$ for $e^+$ when 
the decay channel to pions is not open.

\section{Photon spectrum}
\subsection{Galactic gamma-ray emission}
Pions are produced at rest just after they are kinematically allowed, and immediately they emit photons which
has peculiar spectrum as explained in the following. $\pi^0$ decays into two photons with the energy of half the 
$\pi^0$ mass, 67.5~MeV. We first consider the Galactic gamma-ray emission. The spectrum of this line gamma 
rays can be obtained as
\begin{equation}
\varepsilon_\gamma^2\frac{d\Phi_{2\gamma}}{d\varepsilon_\gamma}=
\frac{1}{4\pi}\frac{{\Omega_Q\cal D}}{\omega_Q \Delta\tau_Q}\frac{Q_\pi}{Q_{\rm D}}f_{\pi^0}
2 \left(\frac{\varepsilon_\gamma}{\Delta\varepsilon_\gamma}\right) \varepsilon_\gamma,
\label{pi02gamma}
\end{equation}
where $\Omega_Q$ is the density parameter of the Q ball. $\Delta\tau_Q=\xi t_0$, where $\xi$ represents 
the uncertainty of the lifetimes of the Q balls due to some variance in their sizes. 
$\Delta\varepsilon_\gamma/ \varepsilon_\gamma$ is the energy resolution of the observation. $Q_\pi/Q_{\rm D}$ 
implies the fraction of the charge of the Q ball that can decay into pions, and $f_{\pi^0}$ is the number of $\pi^0$ 
produced per unit Q-ball charge, as described in the previous section. ${\cal D}$ is a D-factor, given by
\begin{equation}
{\cal D}=\frac{1}{\Delta\Omega} \int_{\Delta\Omega} d\ell db\cos b \int_0^{s_{\rm max}} ds
\rho_{\rm DM} (r(\ell,b,s)),
\end{equation}
where $\ell$ and $b$ are Galactic longitude and latitude, respectively. $\Delta\Omega$ denotes the solid angle 
of the observation. $s$ is the distance along the line of sight. The distance from the Galactic Center is expressed 
as $r(\ell,b,s)=\left( R_{GC}^2 - 2 s R_{\rm GC} \cos\psi +s^2 \right)^{1/2}$, where $\psi$ is the angle between 
the direction to the Galactic Center and the line of sight so that $\cos\psi = \cos\ell\cos b$. 
$s_{\rm max}=\left(R_{\rm MW}^2-\sin^2\psi R_{\rm GC}^2\right)^{1/2}+R_{\rm GC}\cos\psi$ with $R_{MW}=30$~kpc
being assumed as the size of the dark halo. We adopt the Navaro-Frenk-White 
profile \cite{Navarro:1995iw} for the dark matter halo density, written as
\begin{equation}
\rho_{\rm DM}(r) = \frac{\rho_0}{\left(\frac{r}{r_0}\right)\left(1+\frac{r}{r_0}\right)^2}.
\end{equation}
Here we take $\rho_0 = 0.53$~GeV/cm$^3$ and $r_0=15$~kpc so that we have the solar local density of
$\rho_{\rm DM}(R_{\rm GC})=0.4$~GeV/cm$^3$, where $R_{\rm GC}=8.23$~kpc is the distance to the Galactic 
Center from the sun \cite{Leung:2022dno}. 

On the other hand, $\pi^+$ and $\pi^-$ decay into $\mu^+$ and $\mu^-$, respectively. Photons are emitted 
through the internal bremsstrahlung with the branching ratio \cite{Bressi:1997gs,Brown:1964zza,Neville:1961zz}
\begin{eqnarray}
\frac{dB_{\pi{\rm IB}}}{d\varepsilon_\gamma} & = & \frac{\alpha}{2\pi}\frac{1}{(1-p)^2} \frac{2}{m_{\pi^\pm}}\frac{1}{x}
\nonumber \\
& & \times \left [\frac{\left\{x^2+4(1-p)(1-x)\right\}(x^2+p-1)}{1-x}\right. \nonumber \\
& & \left. +\left\{2(1-p)(1+p-x)+x^2\right\}\log\frac{1-x}{p}\right], \nonumber \\
& & 
\label{dBpi}
\end{eqnarray}
where $\alpha=1/137$ is the fine structure constant, $p=(m_\mu/m_{\pi^\pm})^2$, and 
$x=2\varepsilon_\gamma/m_{\pi^\pm}$. See Appendix~A for the details. The spectrum of IB photons is thus given by
\begin{equation}
\varepsilon_\gamma^2\frac{d\Phi_{\pi{\rm IB}}}{d\varepsilon_\gamma}=
\frac{1}{4\pi}\frac{\Omega_Q{\cal D}}{\omega_Q \Delta\tau_Q}\frac{Q_\pi}{Q_{\rm D}}f_{\pi^\pm}
\frac{dB_{\pi{\rm IB}}}{d\varepsilon_\gamma} \varepsilon_\gamma^2,
\label{piIB}
\end{equation}
where $f_{\pi^\pm}=f_{\pi^+}+f_{\pi^-}=7/60$ since the spectrum of internal bremsstrahlung is the same for both
$\pi^+$ and $\pi^-$ decays.

In addition, $\mu^+$ and $\mu^-$ decay into $e^+$ and $e^-$, respectively,  similarly emitting  
internal bremsstrahlung with the branching ratio \cite{Kuno:1999jp}
\begin{eqnarray}
\frac{dB_{\mu{\rm IB}}}{d\varepsilon_\gamma} & = &\frac{\alpha}{2\pi} \frac{2}{m_\mu} \frac{2(1-y)}{3y} \nonumber \\
& & \times \left[-\frac{17}{2}-\frac{3}{4}(1-y)+\frac{16}{3}(1-y)^2-\frac{55}{12}(1-y)^3 \right. \nonumber \\
& & \left. +\left\{3-2(1-y)^2+2(1-y)^3\right\}\log\frac{1-y}{q} \right],
\label{dBmu}
\end{eqnarray}
where $q=(m_e/m_\mu)^2$ and $y=2\varepsilon_\gamma/m_\mu$. See Appendix~A for the details. The spectrum of
the IB photons is then written as
\begin{equation}
\varepsilon_\gamma^2\frac{d\Phi_{\mu{\rm IB}}}{d\varepsilon_\gamma}=
\frac{1}{4\pi}\frac{\Omega_Q{\cal D}}{\omega_Q  \Delta\tau_Q}\frac{Q_\pi}{Q_{\rm D}}f_{\mu^\pm}
\frac{dB_{\mu{\rm IB}}}{d\varepsilon_\gamma} \varepsilon_\gamma^2,
\label{muIB}
\end{equation}
where $f_{\mu^\pm}=f_{\pi^\pm}+f_{\mu^-}=2/15$, since the charged pions decay into muons. 

\begin{figure}[ht!]
\includegraphics[width=90mm]{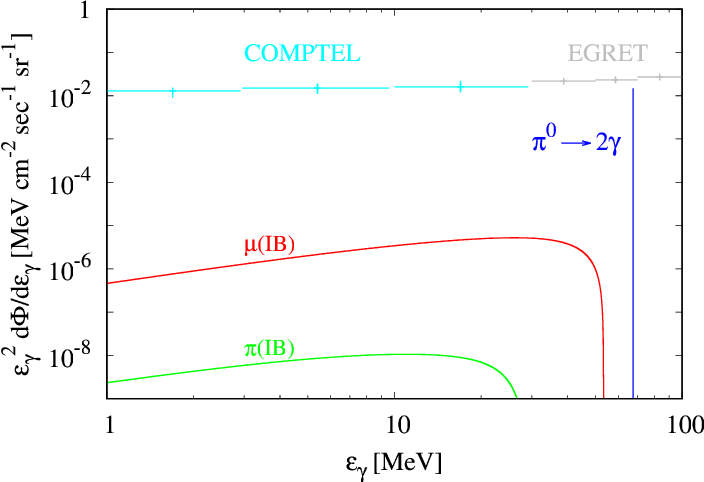} 
\caption{Photon spectra from the Q-ball decay. We display the spectra (\ref{pi02gamma}), (\ref{piIB}) 
and (\ref{muIB}) in blue, green, and red lines, respectively. Observations of the Galactic diffuse 
gamma-ray emission are also shown in cyan (COMPTEL) and gray points (EGRET). 
\label{fig2}}
\end{figure}

We illustrate the spectra (\ref{pi02gamma}), (\ref{piIB}) and (\ref{muIB}) in Fig.~\ref{fig2}. Here we consider 
the region with $|\ell|<30^\circ$ and $|b|<5^\circ$, and set $\Omega_Q/\xi \simeq 1\times 10^{-8}$, well below
the rough upper bound explained in Appendix~B.\footnote{
Since Q-ball abundance would be by far much larger than this value for usual formation \`a la Affleck-Dine, 
it is necessary to reduce it. One mechanism is considered in Ref.~\cite{Kasuya:2005ay} using
resonant decay of the flat direction in most of the region of the universe, and another one may be late-time
entropy production before nucleosynthesis. 
}
Also shown are the observations of the gamma-ray diffuse emission from the inner Galactic region measured 
by COMPTEL in 1--30~MeV \cite{Strong:1998ck} and EGRET in 30--100~MeV \cite{Strong:2004de}.  
We adopt $\Delta\varepsilon_\gamma/\varepsilon_\gamma=0.15$ in Eq.(\ref{pi02gamma}) to compare with 
the EGRET data \cite{Thompson:2008rw}.
It should be noted that the spectra obtained by COMPTEL and EGRET include the Galactic diffuse emission, 
and there may be an excess component in addition to the Galactic diffuse emission to reconcile with the 
observed spectra (See, e.g., Refs.~\cite{Strong:2004de,Fermi-LAT:2012edv,Orlando:2017mvd}). The excess 
could be contributed from the gamma rays originated from the Q-ball decay, unresolved sources \cite{Tsuji:2022pfw}, 
dark matter \cite{Boddy:2015efa,Binder:2022pmf,Christy:2022pvy} and/or cosmic neutrinos \cite{Fang:2022trf},
or may be explained by the uncertainties of the model of the Galactic diffuse emission (e.g., Ref.~\cite{Bouchet:2011fn}).

We can see that the unique feature of the spectrum is the line-gamma rays at $\varepsilon_\gamma=67.5$~MeV. 
Astrophysical sources could hardly produce such an emission line. It could be detected by future observations such as
AMEGO-X \cite{Fleischhack:2021mhc}, CubeSat for MeV observations (MeVCube) \cite{Lucchetta:2022nrm}, 
e-ASTROGAM \cite{e-ASTROGAM:2017pxr}, GECCO \cite{Moiseev:2021mdo}, GRAINE \cite{Rokujo:2018ggw}, 
GRAMS \cite{Aramaki:2019bpi}, and SMILE-3 \cite{Takada:2021iug}.\footnote{
Although the gamma-ray data by Fermi-LAT below 100~MeV is available, there is no dedicated search 
for 67.5~MeV line emission because of the technical difficulty at the lower energy band.}%
\footnote{
COSI-SMEX \cite{Tomsick:2019wvo}, to be launched in 2027, is not sensitive to the 
gamma rays above 5 MeV.}
For example, GRAMS has 67.5 MeV line sensitivity of $\sim10^{-6}$~MeV/cm$^2$/s \cite{Aramaki:2019bpi}.
Detection of this line by some future observations would prove this Q-ball decay scenario.
If not, it may result in lowering the upper bound on $\Omega_Q$. In addition, the Q-ball decay produces
almost the same amount of $\pi^0$ and $\pi^\pm$, so that photons from the radiative muon decay through
the internal bremsstrahlung will appear in the energy range $\varepsilon_\gamma \lesssim 50$~MeV 
with a sharp cutoff at around 50~MeV. Continuum sensitivity of GRAMS may reach to observe this
internal bremsstrahlung on the top of the diffuse background for a satellite mission \cite{Aramaki:2019bpi}. 
Measurements of the 67.5~MeV line and/or this broad spectrum with a cutoff at around 50~MeV could be
the evidence of the Q-ball decay model.

\subsection{Extragalactic gamma-ray emission}
Let us now study the extragalactic gamma-ray emission. For $\pi^0\rightarrow 2\gamma$, taking into account
the cosmic expansion, we can estimate the present flux as \cite{Kawasaki:1997ah,Asaka:1997rv}
\begin{eqnarray}
\varepsilon_\gamma^2\frac{d\Phi^{\rm ext}_{2\gamma}}{d\varepsilon_\gamma} & = & \frac{\varepsilon_\gamma^2}{4\pi} 
\int_{(1-\xi)t_0}^{t_0} dt' \frac{\Omega_Q\rho_{c0}}{\omega_Q \Delta\tau_Q} \frac{Q_\pi}{Q_{\rm D}} f_{\pi^0} \nonumber \\
& & \times 2(1+z)\delta\left[\varepsilon_\gamma'-\frac{m_{\pi^0}}{2}\right],
\end{eqnarray}
where $\rho_{c0}$ is the present critical density of the universe, $z$ is the redshift, and 
$\varepsilon_\gamma'=(1+z)\varepsilon_\gamma$ is the photon energy produced at the decay time for
the photon with energy $\varepsilon_\gamma$ at present. The redshift $z$ is related to cosmic time $t$ as
\begin{equation}
\frac{dt}{dz}=-(1+z)^{-5/2}H_0^{-1}\left[\Omega_{\rm M} +\frac{\Omega_\Lambda}{(1+z)^3}\right]^{-1/2},
\end{equation}
where $H_0$ is the Hubble parameter at present, $\Omega_M$ and $\Omega_\Lambda$ are 
the present density parameters of matter and cosmological constant, respectively, and the flat universe is
assumed. Therefore, we obtain the flux as
\begin{eqnarray}
\varepsilon_\gamma^2\frac{d\Phi_{2\gamma}^{\rm ext}}{d\varepsilon_\gamma} & = & \frac{1}{2\pi} 
\frac{\Omega_Q\rho_{c0}}{\omega_Q \Delta\tau_Q H_0} \frac{Q_\pi}{Q_{\rm D}} f_{\pi^0}
\varepsilon_\gamma \left(\frac{2\varepsilon_\gamma}{m_{\pi^0}}\right)^{3/2}  \nonumber \\
& & \times \left[\Omega_{\rm M}+\Omega_\Lambda \left(\frac{2\varepsilon_\gamma}{m_{\pi^0}}\right)^3 \right]^{-1/2}.
\label{ext2gamma}
\end{eqnarray}
Notice that the spectrum has a rather narrow shape where the upper and lower edges are respectively located at
the energy of 67.5~MeV and $67.5/(1+z_\xi)$~MeV, where $z_\xi$ corresponds to the redshift at the cosmic time $(1-\xi)t_0$.

For the internal bremsstrahlung from pions, we have
\begin{eqnarray}
\varepsilon_\gamma^2\frac{d\Phi_{\pi{\rm IB}}^{\rm ext}}{d\varepsilon_\gamma} & = & \frac{\varepsilon_\gamma^2}{4\pi} 
\int_{(1-\xi)t_0}^{t_0} dt' \frac{\Omega_Q\rho_{c0}}{\omega_Q\Delta\tau_Q} \frac{Q_\pi}{Q_{\rm D}} f_{\pi^\pm} \nonumber \\
& & \times (1+z) \frac{dB_{\pi{\rm IB}}}{d\varepsilon'_\gamma}(\varepsilon'_\gamma),
\end{eqnarray}
where Eq.(\ref{dBpi}) is used. Changing the variable of integration to $z$, we obtain
\begin{eqnarray}
& & \varepsilon_\gamma^2\frac{d\Phi_{\pi{\rm IB}}^{\rm ext}}{\varepsilon_\gamma} =  \frac{\varepsilon_\gamma}{4\pi} 
\frac{\Omega_Q\rho_{c0}}{\omega_Q\Delta\tau_Q H_0} \frac{Q_\pi}{Q_{\rm D}} f_{\pi^\pm} 
\frac{\alpha}{2\pi} \frac{1}{(1-p)^2} \hspace{10mm} \nonumber \\
& & \hspace{15mm}
\times \int_0^{z_\xi} dz \frac{K_\pi \left((1+z)\frac{2\varepsilon_\gamma}{m_{\pi^\pm}}\right)}
{\sqrt{\Omega_{\rm M}(1+z)^5+\Omega_\Lambda(1+z)^2}},
\label{extpiIB}
\end{eqnarray}
where 
\begin{eqnarray}
K_\pi(x) & = & \frac{\left\{x^2+4(1-p)(1-x)\right\}(x^2+p-1)}{1-x} \nonumber \\
& &  +\left\{2(1-p)(1+p-x)+x^2\right\}\log\frac{1-x}{p}. \nonumber \\
& &
\end{eqnarray}

On the other hand, the spectrum of the internal bremsstrahlung from muons can be calculated as
\begin{eqnarray}
\varepsilon_\gamma^2\frac{d\Phi_{\mu{\rm IB}}^{\rm ext}}{d\varepsilon_\gamma} & = & \frac{\varepsilon_\gamma^2}{4\pi} 
\int_{(1-\xi)t_0}^{t_0} dt' \frac{\Omega_Q\rho_{c0}}{\omega_Q\Delta\tau_Q} \frac{Q_\pi}{Q_{\rm D}} f_{\mu^\pm} \nonumber \\
& & \times (1+z) \frac{dB_{\mu{\rm IB}}}{d\varepsilon'_\gamma}(\varepsilon'_\gamma),
\end{eqnarray}
where Eq.(\ref{dBmu}) is exploited. Converting the variable of integration to $z$, we get
\begin{eqnarray}
& & \varepsilon_\gamma^2\frac{d\Phi_{\mu{\rm IB}}^{\rm ext}}{d\varepsilon_\gamma} =  \frac{\varepsilon_\gamma}{4\pi} 
\frac{\Omega_Q\rho_{c0}}{\omega_Q\Delta\tau_Q H_0} \frac{Q_\pi}{Q_{\rm D}} f_{\mu^\pm}
\hspace{25mm} \nonumber \\
& & \hspace{5mm}
\times \int_0^{z_\xi} dz \frac{2\left[ J_+\left((1+z)\frac{2\varepsilon_\gamma}{m_\mu}\right)+J_
+\left((1+z)\frac{2\varepsilon_\gamma}{m_\mu}\right)\right]}
{\sqrt{\Omega_{\rm M}(1+z)^5+\Omega_\Lambda(1+z)^2}}, \nonumber \\
& &
\label{extmuIB}
\end{eqnarray}
where Eqs.(\ref{Jp}) and (\ref{Jm}) are used.

\begin{figure}[ht!]
\includegraphics[width=90mm]{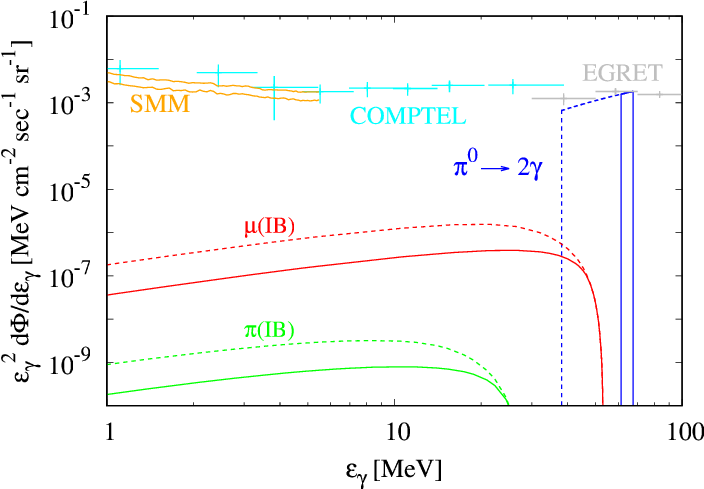} 
\caption{Extragalactic photon spectra from the Q-ball decay. We display the spectra (\ref{ext2gamma}) in blue, 
(\ref{extpiIB}) in green, and (\ref{extmuIB}) in red lines. Solid and dashed lines denote the cases with $\xi=0.1$ 
and $0.5$, respectively. Observations of the extragalactic 
diffuse gamma-ray emission are also shown in cyan (COMPTEL) and gray (EGRET) points, and in orange lines (SMM).
\label{fig3}}
\end{figure}

We plot the spectra of (\ref{ext2gamma}),  (\ref{extpiIB}) and (\ref{extmuIB}) in Fig.~\ref{fig3}. The integrals in 
Eqs.(\ref{extpiIB}) and (\ref{extmuIB}) are estimated numerically. Cosmological parameters are set to be 
$\Omega_M=0.311$, $\Omega_\Lambda=0.689$, and $H_0=67.7$~km/s/Mpc \cite{Planck:2018vyg}. 
We illustrate the cases for $\xi=0.1$ and $0.5$, which correspond respectively to $z_\xi\simeq 0.1$ and $0.8$. 
Also shown are the observations of the extragalactic diffuse gamma-ray emission by EGRET (above 30~MeV) 
\cite{Strong:2004ry}, COMPTEL (below 30~MeV) \cite{Weidenspointer2000}, and the Solar Maximum Mission (SMM) 
Gamma-Ray Spectrometer in the energy range 0.3--8.0~MeV \cite{Watanabe2000}. The origin of this extragalactic 
MeV gamma-ray background remains unknown, and contributions from seyfert galaxies or blazars are suggested 
(e.g., Refs.~\cite{Inoue:2007tn,Ajello:2009ip}).

We can see a thick line-like emission at  $\lesssim 67.5$~MeV,  and broad spectra of internal bremsstrahlung 
with sharp cutoffs at 50~MeV and 30~MeV for muon and pion decays, respectively. They have similar features as in the 
Galactic diffuse gamma-ray spectrum. Future MeV gamma-ray observations could prove or falsify these signals from 
the Q-ball decay.

\section{Conclusion}
We have investigated the observational clues of the SUSY Q balls which decay at present.
Such Q balls are realized by having very large lepton numbers as the charge of the Q ball in the
gauge-mediated SUSY breaking models. The new-type Q balls form and transform to 
the gauge-mediation type towards the end of the decay. Then the mass per unit charge $\omega_Q$ 
increases as the charge $Q$ decreases, and subsequently the decay channels into
pions become kinematically allowed to create pions at rest. This is the distinctive mechanism of the
Q-ball decay to produce pions at rest. 

Neutral pions decay into two photons with the energy of half the pion mass, making the sharp emission 
line at 67.5~MeV. If detected, it would be a smoking gun of the SUSY Q ball since it is the unique feature 
of the Q-ball decay and no astrophysical sources would create such an emission line. In addition, 
charged pions decay into muons which further decay radiatively to emit internal bremsstrahlung whose 
spectrum has a sharp edge at around 50~MeV. We may expect that these photons could be detected 
by future MeV gamma-ray observations. 

\section*{Acknowledgments} 
The authors are grateful to Tsuguo Aramaki for useful comments. This work is supported by JSPS KAKENHI 
Grant Nos. 20H05851(M. K.), 21K03567(M. K.), and 22K14064 (N. T.).

\appendix
\section{Internal bremsstrahlung}
When charged pions decay into muons, they partially emit internal bremsstrahlung. Its differential branching 
ratio is written as  \cite{Bressi:1997gs,Brown:1964zza,Neville:1961zz}
\begin{eqnarray}
\frac{dB_{\pi{\rm IB}}}{dxdw} & = & \frac{\alpha}{2\pi}\frac{1}{(1-p)^2}\frac{1-w+p}{x^2(x+w-1-p)} \nonumber \\
& & \times \left[ x^2+2(1-x)(1-p) -\frac{2xp(1-p)}{x+w-1-p}\right], \nonumber \\
& &
\end{eqnarray}
where $\alpha=1/137$ is the fine structure constant, $p=(m_\mu/m_{\pi^\pm})^2$, 
$x=2\varepsilon_\gamma/m_{\pi^\pm}$, and $w=2E_\mu/m_{\pi^\pm}$ is the normalized muon energy.
$w$- and $x$-ranges are
\begin{equation}
\begin{array}{rcl}
2\sqrt{p} \le & w & \le 1+p, \\[2mm]
1-\frac{w}{2}-\frac{\sqrt{w^2-4p}}{2} \le & x & \le 1-\frac{w}{2}+\frac{\sqrt{w^2-4p}}{2},
\end{array}
\label{wrange}
\end{equation}
respectively. Integration with respective to $w$ over the $w$-range of (\ref{wrange}) leads to
\begin{eqnarray}
\frac{dB_{\pi{\rm IB}}}{dx} & = & \frac{\alpha}{2\pi}\frac{1}{(1-p)^2} \frac{1}{x}
\nonumber \\
& & \times \left [\frac{\left\{x^2+4(1-p)(1-x)\right\}(x+p-1)}{1-x}\right. \nonumber \\
& & \left. +\left\{2(1-p)(1+p-x)+x^2\right\}\log\frac{1-x}{p}\right]. \nonumber \\
& & 
\end{eqnarray}

On the other hand, polarized muons are produced by the decay of charged pions at rest. The photon
spectrum by radiative muon decay is given by \cite{Kuno:1999jp}
\begin{eqnarray}
\frac{dB(\mu^\pm\rightarrow e^\pm\nu\bar{\nu}\gamma)}{dyd\cos\theta_\gamma} & = &
\frac{1}{y}\left[ J_+(y) \left(1\pm P_\mu\cos\theta_\gamma \right) \right. \nonumber \\
& & \left. + J_-(y) \left(1\mp P_\mu\cos\theta_\gamma \right) \right],
\label{dBdydcostheta}
\end{eqnarray}
where $y=2\varepsilon_\gamma/m_\mu$, $P_\mu$ is a magnitude of the muon spin polarization vector,
$\theta_\gamma$ is the angle between the muon spin polarization and the photon momentum.
$J_+(y)$ and $J_-(y)$ are written respectively as
\begin{eqnarray}
J_+(y) & = & \frac{\alpha}{6\pi}(1-y)\left[ \left(3\log\frac{1-y}{q}-\frac{17}{2}\right)\right. \nonumber \\
& & +\left(-3\log\frac{1-y}{q}+7\right)(1-y) \nonumber \\
& & \left.+\left( 2\log\frac{1-y}{q}-\frac{13}{3}\right)(1-y)^2\right],
\label{Jp}
\end{eqnarray}
\begin{eqnarray}
J_-(y) & = & \frac{\alpha}{6\pi}(1-y)^2\left[ \left(3\log\frac{1-y}{q}-\frac{31}{4}\right)\right. \nonumber \\
& & +\left(-4\log\frac{1-y}{q}+\frac{29}{3}\right)(1-y) \nonumber \\
& & \left.+\left( 2\log\frac{1-y}{q}-\frac{55}{12}\right)(1-y)^2\right],
\label{Jm}
\end{eqnarray}
where $q=(m_e/m_\mu)^2$. Since the produced muons have no specific direction, we must integrate 
Eq.(\ref{dBdydcostheta}) with respect to $\theta_\gamma$ over the whole angles, and hence obtain the spectrum as
\begin{eqnarray}
\frac{dB_{\mu{\rm IB}}}{dy} & = & \frac{2}{y}\left[ J_+(y)  + J_-(y)  \right] \nonumber \\
& = & \frac{\alpha}{2\pi} \frac{2(1-y)}{3y} \nonumber \\
& & \times \left[-\frac{17}{2}-\frac{3}{4}(1-y)+\frac{16}{3}(1-y)^2-\frac{55}{12}(1-y)^3 \right. \nonumber \\
& & \left. +\left\{3-2(1-y)^2+2(1-y)^3\right\}\log\frac{1-y}{q} \right].
\end{eqnarray}

\section{Rough constraint on $\Omega_Q$}
Q-ball decay  produces positrons with energy $\lesssim \omega_Q$ which may annihilate with electrons
at the Galactic Center to generate 511~keV gamma rays. Since the morphology of the dark matter halo 
in our galaxy is still not known, we make very rough estimate for the upper bound of the Q-ball abundance
according to Ref.~\cite{Hooper:2004qf}. Assuming the half of the total 511~keV flux is 
emitted from an angular region of $9^\circ$ circle, we have
\begin{equation}
\frac{M_{(<9^\circ)}\Omega_Q}{\omega_Q \Delta\tau_Q} f_{e^+} \left[\frac{f}{4}+(1-f)\right]
=\frac{1}{2} \Phi_{511} 4\pi R_{\rm GC}^2,
\end{equation}
where $f$ is the fraction of positrons which annihilate via positronium, $\Phi_{511}(\simeq 10^{-3}$~cm$^{-2}$s$^{-1}$)
is the observed total flux of 511~keV line, and $\Delta\tau_Q=\xi\tau_Q$. We assume $\tau_Q \simeq t_0$. The total 
mass within the $9^\circ$ circle is given by
\begin{equation}
M_{(<9^\circ)} = \int_{(<9^\circ)} \rho_{\rm DM}(r) 4\pi r^2 dr.
\end{equation}
Therefore, we obtain the constraint on the density parameter as
\begin{eqnarray}
\frac{\Omega_Q}{\xi} & \lesssim & 2.4\times 10^{-7} 
\left(\frac{\Phi_{511}}{10^{-3}~{\rm cm}^{-2}{\rm s}^{-1}}\right)  \nonumber \\
& & \times 
\left(\frac{R_{\rm GC}}{8.23~{\rm kpc}}\right)^2 
\left(\frac{\omega_Q}{2.9 {\rm MeV}}\right)\left(\frac{\tau_Q}{t_0}\right).
\end{eqnarray}



\end{document}